# Tetragonal BaCoO$_3$: A Co$^{4+}$ Ferromagnetic Mott Insulator with Inverted Spin Crossover

Mingyu Xu[1], Haozhe Wang[1], Krishna Prasad Koirala[2], Corey Melnick[3], Cheng Peng[1], Mario U. González-Rivas[4,5], Jiaqi Lu[6], Le Wang[2], Mark H. Engelhard[7], Yingge Du[2], Xianglin Ke[8], Robert J. Green[5,9], Alannah M. Hallas[4,5], Jie Li[6], Gabriel Kotliar[3,10], Weiwei Xie[1*]

1. Department of Chemistry, Michigan State University, East Lansing, MI 48824, USA
2. Physical and Computational Sciences Directorate, Pacific Northwest National Laboratory, Richland, WA 99354, USA
3. Condensed Matter Physics and Materials Science Department, Brookhaven National Laboratory, Upton, NY 11973, USA
4. Department of Physics & Astronomy, University of British Columbia, Vancouver, BC V6T 1Z1, Canada
5. Stewart Blusson Quantum Matter Institute, University of British Columbia, Vancouver, BC V6T 1Z4, Canada
6. Department of Earth and Environmental Sciences, University of Michigan, Ann Arbor, MI 48109, USA
7. Energy and Environment Directorate, Pacific Northwest National Laboratory, Richland, WA 99354, USA
8. Department of Physics and Astronomy, Michigan State University, East Lansing, MI 48824, USA
9. Department of Physics and Engineering Physics, University of Saskatchewan, Saskatoon S7N 5E2, Saskatchewan, Canada
10. Department of Physics and Astronomy, Rutgers University, Piscataway, NJ 08854, USA

[*]Corresponding authors: Dr. Weiwei Xie (xieweiwe@msu.edu)

## *Abstract*

The interplay between crystal electric field splitting of *d* states and Hund's rule exchange energy in cobalt-based perovskites offers a promising avenue for inducing spin-state transitions. This study reports a new body-centered tetragonal (BCT) phase of $BaCoO_3$ (BCT-$BaCoO_3$), synthesized under high pressure (15 GPa) and high temperature (1200 °C) conditions. BCT-$BaCoO_3$ adopts a double perovskite structure of $EuTiO_3$-type (space group *I*4/*mcm*, #140), confirmed by high-resolution scanning transmission electron microscopy. X-ray photoelectron spectroscopy reveals a rare $Co^{4+}$ valence state. Magnetization and X-ray absorption measurements reveal a low-spin to high-spin transition that takes place between 200 and 300 K. While spin crossovers are relatively common among common oxides, the one observed in BCT-$BaCoO_3$ is remarkable in that it proceeds in the opposite direction from conventional spin transitions. BCT-$BaCoO_3$ exhibits a low-spin ($S = ½$) state at high temperatures and transitions to a high-spin ($S = 5/2$) state at low temperatures. Within the high-spin state, hard ferromagnetic order onsets at $T_C = 107$ K. Electrical resistivity indicates weak magnetoresistance and insulating behavior. Overall, BCT-$BaCoO_3$ presents an exceptional model for the exploration of spin-state transitions and the study of Co spin states in cobalt-based perovskites.

## Introduction

Transition metal oxides with perovskite crystal structures exhibit a broad spectrum of magnetic and electronic phenomena, including high-temperature superconductivity[1,2], colossal magnetoresistance[3], ferroelectricity[4], as well as metal-to-insulator, structural, and magnetic phase transitions. These perovskites, with the general formula $ABO_3$, can adopt either a three-dimensional framework of corner-sharing $[BO_6]$ octahedra (in cubic (C) or pseudo cubic forms) or one-dimensional chains of face-sharing $[BO_6]$ octahedra in hexagonal (H) arrangements.[5] Cobalt-based perovskites (B = Co), such as the hexagonal 2H-$BaCoO_3$[6,7] and cubic $SrCoO_3$[8], exemplify this diversity. In 2H-$BaCoO_3$, face-sharing $[CoO_6]$ octahedra form parallel chains along the *c*-axis, creating a two-dimensional triangular lattice in the *ab*-plane with a $Co^{4+}$–$O^{2-}$–$Co^{4+}$ angle of approximately 78°.[7] This structure contrasts with $SrCoO_3$, where the $Co^{4+}$ ions are linked in three equivalent directions by 180° $Co^{4+}$–$O^{2-}$–$Co^{4+}$.[8] Substituting $Sr^{2+}$ and $Ba^{2+}$ with smaller $Ca^{2+}$ ions lead to $Ca_3Co_2O_6$ in the $Sr_4PtO_6$-type structure, diverging from the typical $ABO_3$ structure.[9] The orthorhombic $GdFeO_3$-type structure of $CaCoO_3$ is stabilized exclusively through high-pressure oxygen annealing,[10] showcasing the structural complexity of these materials.

Cobalt-based perovskites have garnered significant interest for their spin-state transitions and unusual Co spin states observable within experimentally accessible temperature ranges.[11] This is attributed to the comparable magnitudes of crystal electric field (CEF) splitting of Co *d* states and Hund's rule exchange energy.[12] The small energy gap between the $t_{2g}$ and $e_g$ states enables the thermal excitation of $t_{2g}$ electrons to $e_g$ states, resulting in higher spin states.[13,14] Spin-state transitions have been reported in $LaCoO_3$[15,16], $Pr_{1-x}Ca_xCoO_3$[17], and $GdBaCo_2O_{5.5}$[18], with intermediate spin states observed in $LaCoO_3$[19] and $La_{1-x}Sr_xCoO_3$[20,21]. Conversely, 2H-$BaCoO_3$ exhibits weak magnetism, with its high-temperature magnetic susceptibility suggesting the presence of low-spin $S = 1/2$ $Co^{4+}$. In contrast, $SrCoO_3$, a ferromagnetic (FM) metal with a Curie temperature $T_C$ of ~280 K,[22] showcases the complexity due to $Co^{4+}$'s nominally $d^5$ character (yielding spin-1/2 and quantum effects), strong O *p*–Co *d* orbital hybridization, and spin-orbit coupling (SOC).[23] The origin of the near room temperature ferromagnetism in cubic perovskite $SrCoO_3$ has been the subject of extensive discussion.[24] Further reducing the size of A-site cation, causing significant tilting of $[CoO_6]$ octahedra and narrowing of electronic band width, results in CaCoO$_3$ exhibiting antiferromagnetic (AFM) ordering at 95 K with a probable helical spin arrangement, while it remains in an incoherent metallic state even at low temperatures.[25]

The dependence of magnetism on A-site cation size in ACoO$_3$ perovskites inspired our high-pressure investigation of BaCoO$_3$ to tune lattice parameters and crystal structures without introducing new elements. We recently reported that BaCoO$_3$ retains its 2H-type crystal structure with a shorter Co–Co distance of approximately 2.07 Å, synthesized at 6 GPa and 1200 °C for 3 hours.[7] The potential for archiving even shorter Co–Co distances, possibly inducing complex magnetism and unusual Co spin states, motivated further exploration at higher pressures. Here, we discovered that treatment at 15 GPa and 1200 °C for 3 hours transforms 2H-BaCoO$_3$ into a tetragonal symmetry perovskite phase, termed as BCT-BaCoO$_3$. In stark contrast to the cobalt-based perovskites described above, BCT-BaCoO$_3$ undergoes a spin transition that proceeds in the opposite direction to conventional spin crossover transitions, that is electrons are excited into the $e_g$ states at *low* temperature. In addition to this inverse spin crossover, magnetization and electrical resistivity measurements reveal that BCT-BaCoO$_3$ uniquely behaves as a ferromagnetic insulator with $T_C$ of approximately 107 K. This discovery presents exciting opportunities to explore spin-state transitions and uncover unusual Co spin states in Co-based perovskites.

## Results and Discussion

Due to the limited quantity of the sample synthesized under high pressure conditions (15 GPa) using a multi-anvil apparatus, the available material is insufficient to perform X-ray diffraction measurements required for structural determination and phase analysis. We investigated the crystallographic properties of BCT-BaCoO$_3$ using atomically resolved high-angle annular dark field (HAADF) and integrated differential phase contrast (iDPC) imaging techniques in a scanning transmission electron microscope (STEM). **Fig. 1a** presents the body-centered tetragonal perovskite unit cell. **Fig. 1b**, a representative HAADF-STEM image acquired along the *b*-axis, confirms BCT-BaCoO$_3$'s high crystallinity over a large area. **Fig. 1c** presents the Fast Fourier Transform (FFT) of the HAADF-STEM image. The diffraction spot arrangement in the FFT matches well with the tetragonal double perovskite EuTiO$_3$-type structure (*I*4/*mcm*, #140). Notably, we observed an *n*-layer twin feature on the (-202) plane, indicating a superstructure phase. Such twin features typically arise from thermal stress applied during crystal growth.[26] **Fig. 1d** offers atomically resolved HAADF images, clearly visualizing Ba and Co atom positions. **Fig. 1e** highlights and overlays these positions on the images. **Fig. 1f**, an iDPC image, shows the tetragonal structure, including the arrangement of oxygen atoms. Energy dispersive X-ray spectroscopy (EDS) mapping, shown in Extended Data **Fig. 1**, confirmed the uniform distribution of elements in BCT-BaCoO$_3$. EDS analysis, summarizing the concentration variations of Ba, Co, and O atoms, is presented in Extended Data **Fig. 2** and **Table 1**. Significantly, within the margin of error, no detectable oxygen vacancies were observed.

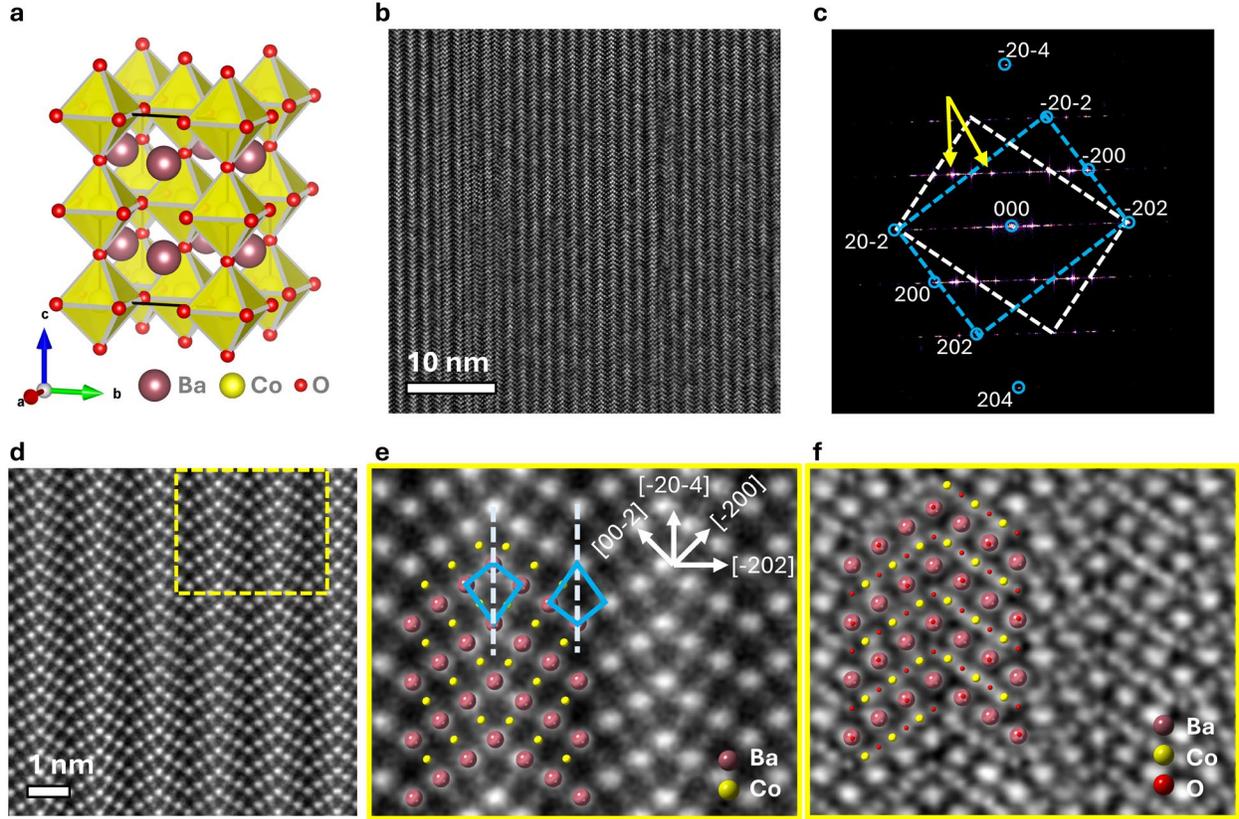

**Fig. 1 | Crystal structure determination of BCT-BaCoO₃. a,** Unit cell of the BCT-BaCoO₃ phase, with Ba, Co, and O represented by brown, yellow, and red, respectively. **b,** HAADF image of an *n*-layer twin structure viewing along the [010] direction. **c,** FFT pattern from the HAADF image; extra spots indicated by yellow arrows result from periodic twins. **d,** Atomically resolved HAADF image depicting the atomic arrangement. **e,** Magnified HAADF image from the yellow box in **d**, overlaid with the tetragonal BCT-BaCoO₃ structure and *n*-layer twin along the (-202) plane. Blue kites highlight the twin boundaries. **f,** iDPC image overlaid with the tetragonal BCT-BaCoO₃ structure, showing the distribution of oxygen atoms.

Fig. 2 compares the unit cells and structures of BCT-BaCoO$_3$, 2H-BaCoO$_3$, SrCoO$_3$, and CaCoO$_3$. In BCT-BaCoO$_3$, corner-sharing [CoO$_6$] octahedra form a 3D network, with Ba atoms occupying cuboctahedra cavities in a perovskite structure. Co–O–Co angles are 180° along the *c*-axis and approximately 173° in the *ab*-plane, attributed to the slight rotation of [CoO$_6$] octahedra parallel to the *c*-axis. The Co–O bond lengths in BCT-BaCoO$_3$ are 1.95 Å, varying slightly, compared to 1.892(1) Å in 2H-BaCoO$_3$. Consequently, [CoO$_6$] octahedra in BCT-BaCoO$_3$ are larger than in the ambient pressure phase 2H-BaCoO$_3$, with the expected increase in density from hexagonal to (pseudo) cubic BaCoO$_3$. The low atomic packing density and close Co–Co distances in 2H-BaCoO$_3$ likely induce the observed structure transition, lowering the system's total energy in response to external pressure.

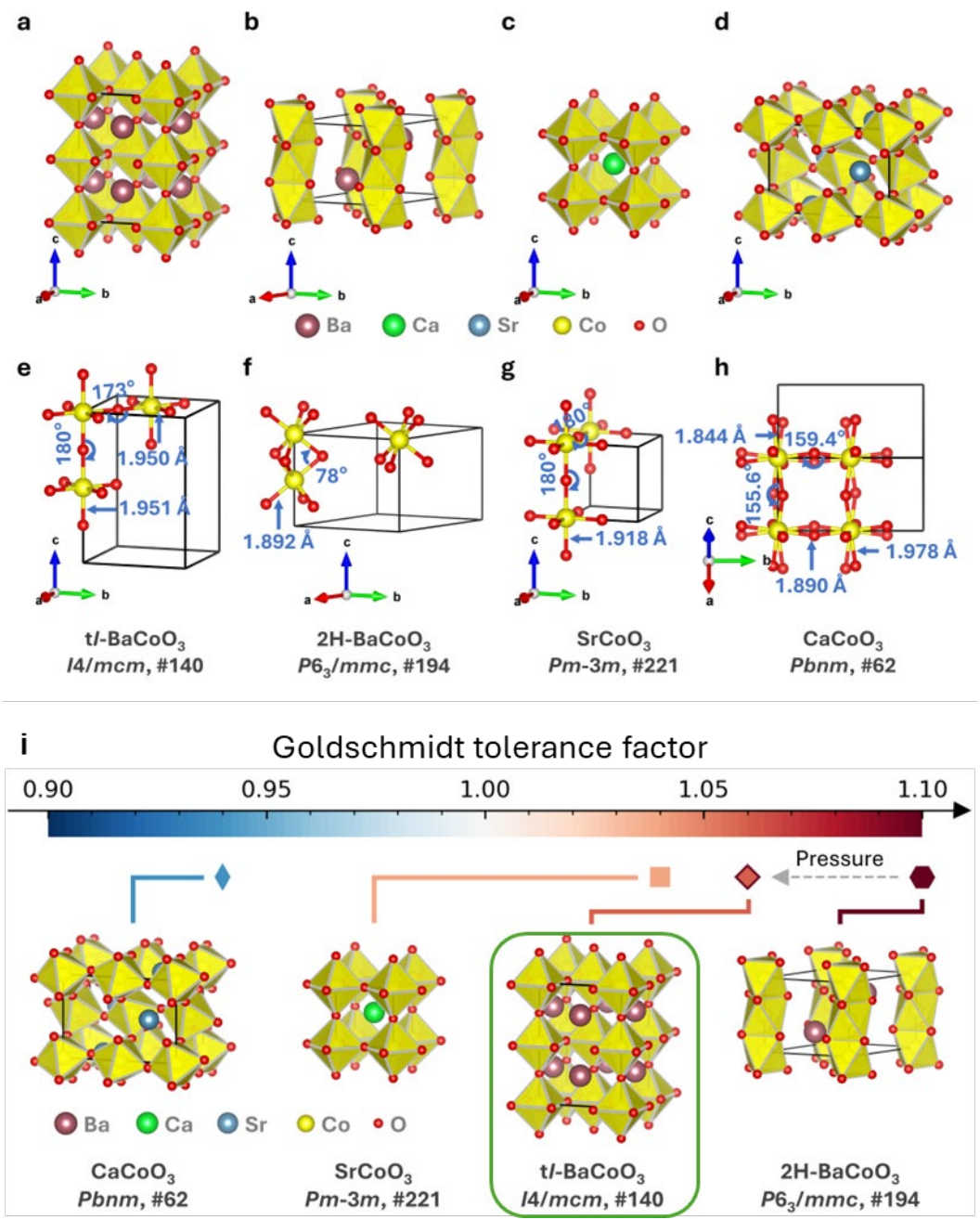

**Fig. 2 | Crystal structure comparisons of BCT-BaCoO₃, 2H-BaCoO₃, SrCoO₃, and CaCoO₃.**
**a–d,** Unit cell of BCT-BaCoO₃ (*I*4/*mcm*, #140), 2H-BaCoO₃ (*P*6₃/*mmc*, #194), SrCoO₃ (*Pm*-3*m*, #221), and CaCoO₃ (*Pbnm*, #62). Brown, green, blue, yellow, and red indicate Ba, Sr, Ca, Co, and O atoms, respectively. **e–h,** Comparison of Co–O bond lengths and Co–O–Co bond angles, with an emphasis on structural anisotropy in these structures. **i,** Goldschmidt tolerance factors and the structural stability of BCT-BaCoO₃, 2H-BaCoO₃, SrCoO₃, and CaCoO₃.

Structure stability and diversity in ACoO$_3$ can be manipulated through Goldschmidt tolerance factor by modifying the atomic size on A-site. Unlike BaCoO$_3$, SrCoO$_3$ with the smaller Sr cation occupying the A-site exhibits a CaTiO$_3$-type cubic structure (*Pm*-3*m*, #221) with regular octahedra forming a 3D network and Co–O bond lengths of 1.918 Å. The replacement of the A-site with an even smaller cation in CaCoO$_3$ results in an orthorhombic GdFeO$_3$-type structure, with similar *a* and *c* axes resulting from [CoO$_6$] octahedra distortion and rotation, leading to Co–O bond lengths of 1.844(1) and 1.978(1) Å. Although the phase was synthesized under high-pressure and high-temperature conditions, the atomic radius correlations still conform to the Goldschmidt tolerance factor rule in **Fig. 2*i***.

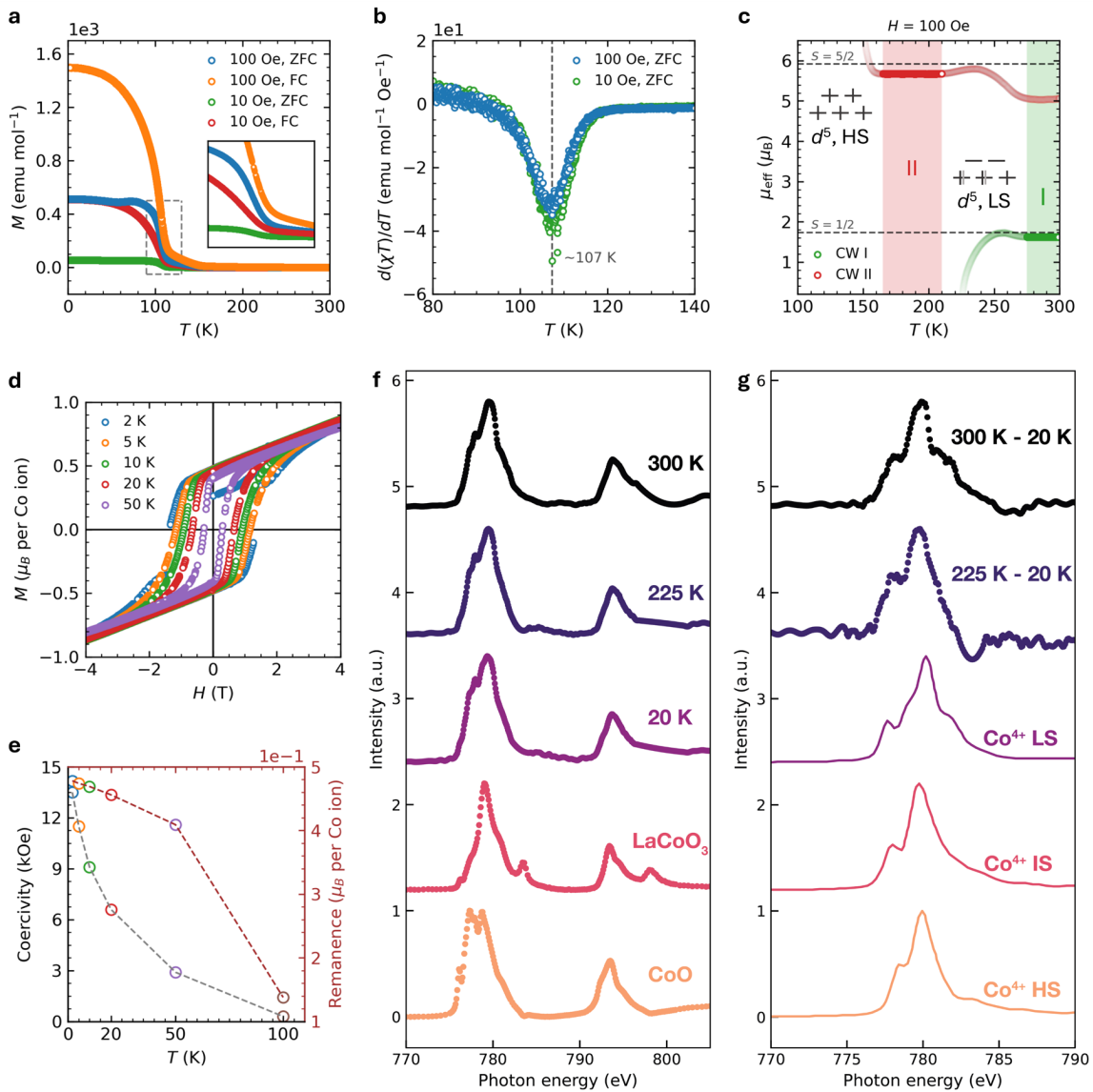

**Fig. 3 | Temperature-dependent magnetization of BCT-BaCoO₃. a,** Magnetic susceptibility as a function of temperature at 10 Oe and 100 Oe, in ZFC and FC modes. Inset: Detailed view of the area within the grey dashed box, near the FM ordering temperature. **b,** Temperature derivative of $\chi T$, indicating the onset of ferromagnetic ordering around 107 K, as suggested by the minimum. **c,** Magnetic susceptibility in temperature Regions I and II, analyzed using the Curie-Weiss law. Regions I and II are depicted in green and red, respectively. **d,** Effective magnetic moments, as determined by the Curie-Weiss law in Region I (green) and II (red). The models extrapolate temperatures beyond their defined regions, thereby providing mutual clarification. Spin-only moments for $S = 1/2$ and $S = 5/2$ states are indicated by dark grey dashed lines. Additionally, schematic diagrams of low-spin (LS) and high-spin (HS) states of the $d^5$ configuration in the standard octahedral ($O_h$) field are included. **e,** Temperature-dependent coercivity and remanence increase as temperature decreases. **f,** XAS spectra for BCT-BaCoO₃ collected at 300 K, 225 K, and 20 K and two standard materials, LaCoO₃ and CoO. **g,** XAS difference spectra (300 K – 20 K and 225 K – 20 K) and Co⁴⁺ multiplet calculations for different possible spin states.

To investigate the magnetic properties of BCT-BaCoO₃, we measured temperature-dependent magnetization in zero-field cooled (ZFC) and field cooled (FC) modes at 10 Oe and 100 Oe, as illustrated in **Fig. 3a**. A ferromagnetic transition was identified at $T_C$ = 107 K, indicated by the sharp increase in the susceptibility while the ordering temperature was defined by taking the minimum in the temperature derivative of $\chi T$, depicted in **Fig. 3b**. Below the FM ordering temperature, the magnetic susceptibility exhibits significant splitting between ZFC and FC modes suggesting a relatively hard ferromagnetic state with significant hysteresis. Compared to ambient pressure 2H-BaCoO₃, BCT-BaCoO₃ shows a marked enhancement in FM interactions, likely due to changes in the Co–O–Co bond angle. This is crucial for determining magnetic super-exchange interactions that significantly affect the magnetic behavior.[27]

Near room temperature, this material predominantly exhibits Curie-Weiss paramagnetism. **Fig. S4** presents the inverse magnetic susceptibility above the FM transition temperature, revealing complex magnetic behavior. Within this temperature range, we identify two distinct temperature regions: 165–210 K (Region I) and 275–300 K (Region II). In Region I, the magnetic susceptibility data were fitted with modified Curie-Weiss law (**Equation 1**),

$$\chi = \chi_0 + \frac{C}{T - \theta_{CW}}$$

where $\theta_{CW}$ is the paramagnetic Curie-Weiss temperature, $\chi_0$ is the temperature independent susceptibility and $C$ is the Curie constant. The fit yielded a $\theta_{CW}$ of 225.6(4) K and a $\mu_{eff}$ of 1.62(1) $\mu_B$ per formula, as detailed in **Fig. S4**. The $\mu_{eff}$ value closely approximates the calculated spin-only moment of 1.73 $\mu_B$ for low-spin (LS) Co$^{4+}$ ($d^5$ electron configuration, $S = 1/2$). In Region II, Curie-Weiss behavior is still observed, shown in **Fig. S4**, providing $\theta_{CW}$ of 97.8(2) K and $\mu_{eff}$ of 5.67(1) $\mu_B$ per formula, in which the effective magnetic moment is much larger than Region I. We note that the spin-only moment for high-spin (HS) Co$^{4+}$ with $S = 5/2$ is 5.92 $\mu_B$. Consequently, a temperature-induced crossover between LS and HS configurations in this system may be attributed to comparable levels of crystal electric field (CEF) splitting ($\Delta_{CEF}$) and Hund's rule exchange energy ($J_{ex}$). Unlike other spin state transitions where LS state could be thermally activated to HS state, this HS-LS crossover indicates that in this system, $\Delta_{CEF}$ is unusually smaller than $J_{ex}$. Together with structure information displayed in **Fig. 1**, we would assign this as a result of longer Co–O bond length in [CoO$_6$] octahedra compared to 2H-BaCoO$_3$, which reduces electron repulsion and then gives a smaller $\Delta_{CEF}$. This crossover is more evident in **Fig. 3c**, where $\mu_{eff}$ values are plotted against temperature using two Curie-Weiss models. In the temperature range between Region I and II, we hypothesize that either an intermediate-spin (IS) state with $S = 3/2$ appears, or the systems behaves as the mixture of LS and HS states, as has been proposed for LaCoO$_3$.[16] These spin states may be accompanied by a second order structural phase transition. However, we assume no tetragonal elongation or compression of [CoO$_6$] octahedra in Region I, with Co $d$ states split in a standard octahedral field, indicating an absence of Jahn-Teller effect (JTE). In fact, for $d^5$ electron configuration, JTE will not create an energy difference for HS state, and only weak JTE is possible for LS state as well as IS state because of the unevenly occupied $t_{2g}$ orbitals. Upon further cooling from Region II, ferromagnetic interactions predominate, leading to a breakdown of Curie-Weiss behavior.

The field-dependent magnetization of BCT-BaCoO$_3$, illustrated in **Fig. 4**, was measured. **Fig. 3d** displays the hysteresis loop observed below the FM ordering temperature, notably below 50 K. Coercivity and remanence increase as the temperature decreases, reaching maximum values of approximately 13.5 kOe and 0.48 $\mu_B$, respectively, at 2 K, as demonstrated in **Fig. 4e**. High coercivity at low temperatures indicates hard ferromagnetism within the system. Magnetization did not reach full saturation up to 7 T, where the magnetic moment measured approximately 1.09 $\mu_B$/Co at 2 K (See **Fig. S5**). This observed moment, suggestive of an unsaturated $S = 5/2$ HS Co$^{4+}$ state, provides further evidence for the LS-HS crossover upon cooling, just above the FM ordering temperature. **Fig. S5** illustrates the field-dependent magnetization above 100 K, with negligible hysteresis observed. Linear responses at 200 K and 300 K confirm the paramagnetic behavior at high temperatures.

X-ray absorption spectroscopy (XAS) measurements allow us directly to probe unoccupied Co $d$ orbitals in BaCoO$_3$ and then determine their spin state. Normalized XAS data for BCT-BaCoO$_3$ and Co$^{2+}$ (CoO) and Co$^{3+}$ (LaCoO$_3$) standards are presented in **Fig. 3f**. No standard is available for Co$^{4+}$. However, we can reliably distinguish it from Co$^{2+}$ and Co$^{3+}$ by the shift in the centroid towards higher energies. Representative data sets of BCT-BaCoO$_3$ are shown above (300 K) and below (225 K) its low spin to high spin crossover and within its magnetically ordered state (20 K). For BCT-BaCoO$_3$, all spectra are mainly assigned to Co$^{4+}$, as is evident from a shift of the spectra' centroids to higher energies compared to the 2+ and 3+ references. However, we also observe fractions of Co$^{2+}$ and Co$^{3+}$, which can be identified from small inflections in the rising edge. The spectra presented here were collected in total electron yield (TEY) mode, which is known to be surface sensitive, suggesting these extra contributions are surface charge states, compensating for the rather unstable ligand-hole nature of the Co$^{4+}$ state. However, the Co$^{4+}$ signals observed here can still be expected to represent the bulk behavior of this species in the sample. Direct assignment of the spin state from the different temperatures' XAS is not possible due to the presence of various contributing signals (Co in 2+, 3+, and 4+ states). However, the change in spin state can still be inferred by monitoring the change in the difference spectra obtained by subtracting the 20 K signal from the higher temperature spectra. These are shown (normalized) in **Fig. 3g** along with reproduced representative ligand field multiplet theory calculations for the low, intermediate, and high spin states of Co$^{4+}$.[30] Most notably, the shoulder at 782 eV on the right-hand side of the central absorption peak in the 300 – 20 K difference trace is in good agreement with only the spectra of Co$^{4+}$ LS, while that shoulder is strongly suppressed in the 225 – 20 K difference trace. The XAS therefore corroborates an electronic transition from Co$^{4+}$ LS to HS.

The electronic properties of cobalt oxides are inherently linked to the electronic configuration of Co$^{4+}$ ions, making the investigation of the temperature-dependent electrical resistivity of BCT-BaCoO$_3$ essential for understanding its behavior. Accordingly, we performed temperature-dependent electrical resistivity measurements from 2 to 300 K under an applied field of up to 5 T on a bulk polycrystalline sample, as shown in **Fig. 4a**. At room temperature, its zero-field resistivity, approximately 25 mΩ cm, is characteristic of poorly conducting oxides.[11] Resistivity increases gradually as temperature decreases to 100 K, then a minor field-dependent feature appears near the FM ordering temperature, shown in **Fig. 4a** inset. Should $Co^{3+}$ ions also exist in the BCT-$BaCoO_3$ phase alongside $Co^{4+}$ ions, Co–O–Co double exchange might induce a metallic state. However, we observed insulating behavior, with resistivity rising from $10^{-1}$ Ω cm at 300 K to nearly $10^5$ Ω cm at 20 K. Furthermore, the observed negligible magnetoresistance (MR) contrasts with the expected significant negative MR in double-exchange ferromagnets.[28] The feature near the FM ordering temperature is more apparent upon examining the temperature derivative of electrical resistivity, as depicted in **Fig. 4b**. This feature appears to be quenched under fields following the FM ordering, indicating a link to changes in the magnetic structure. **Fig. S6** clearly delineates the temperature ranges of various processes, including thermal activation of charge carriers below the FM transition temperature (65–110 K), and Mott variable-range hopping (VRH) in Region I (245–300 K) and II (165–210 K).[29]

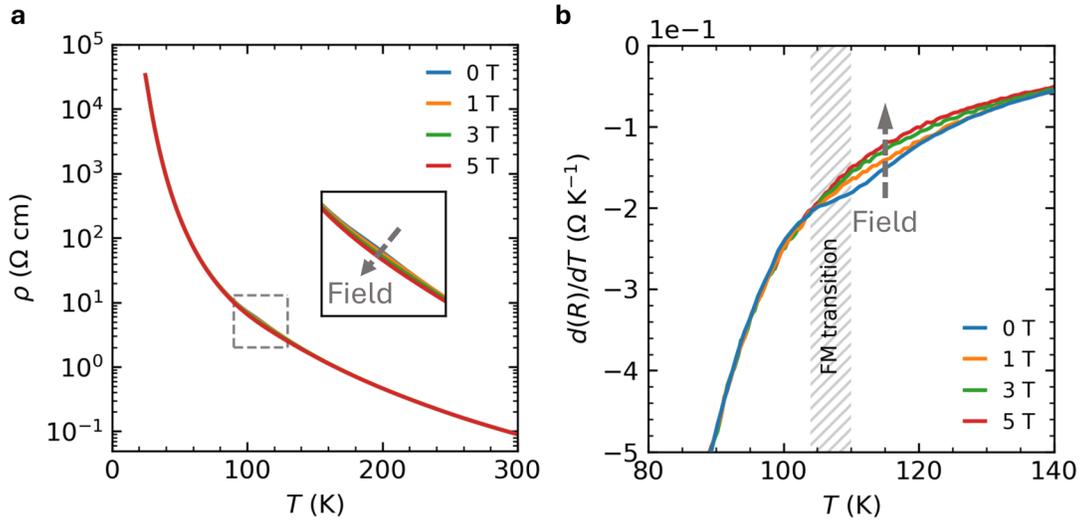

**Fig. 4 | Field and temperature dependence of electrical resistivity in BCT-$BaCoO_3$. a,** Electrical resistivity as a function of temperature under an applied field up to 5 T. Inset: A detailed view within the grey dashed box highlights subtle differences across fields. **b,** Temperature derivative of electrical resistivity, showing a small feature quenched under fields near the FM ordering temperature region.

To elucidate the magnetic and electrical properties of tetragonal BCT-$BaCoO_3$, we performed density functional theory (DFT) calculations. DFT suggests paramagnetic $BaCoO_3$ exhibits metallic behavior with a $d^7$ valence and a Kondo-like peak at the Fermi level, attributable to Co $t_{2g}$ bands, as depicted in **Fig. 5a** and **b**. Incorporating dynamical mean-field theory (DMFT) into DFT calculations, in contrast, reveals fluctuations between the $d^5$ and $d^6$ valences, as illustrated in **Fig. 5c–e**. Additionally, $t_{2g}$ orbitals experience a Mott transition, resulting in the opening of a very small gap (and a larger pseudo-gap) in the spectra, as shown in **Fig. 5c** and **d**. $N = 5$ multiplets show minimal double occupancy, in contrast to $N = 6$ multiplets which necessitate some double occupancy, indicating a negligible preference for $t_{2g}$ or $e_g$ orbital occupation. For $N = 6$ multiplets, a slight preference exists for $e_g$ orbitals due to crystal electric fields. This aligns with our experimental observations. Notably, electrons in both $N = 5$ and $N = 6$ multiplets fluctuate into spin-flipped states, as depicted in **Fig. 5f** and **g**, despite strong Hund's coupling.

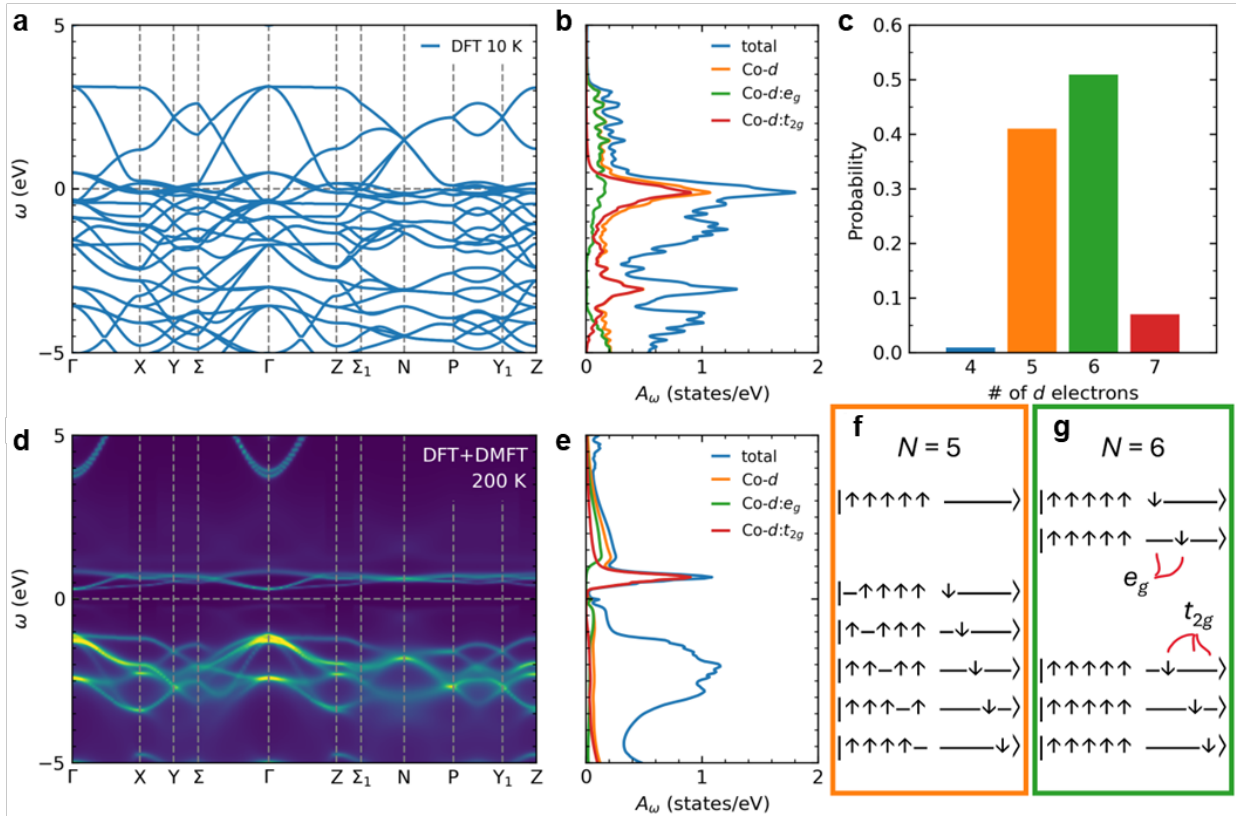

**Fig. 5 | Electronic structure calculations of BCT-BaCoO$_3$. a–d,** Band structure and density of states analyzed using DFT, alongside angle-resolved and integrated spectral functions derived from DFT+DMFT. DFT suggests metallic behavior; conversely, DFT+DMFT reveals a minimal gap, attributed to the divergence of $t_{2g}$ electron self-energy. Incoherent, localized $e_{2g}$ states persist in the pseudo-gap just below the Fermi level, resulting in an enlarged pseudo-gap. **e,** DFT+DMFT impurity histograms show significant fluctuations in the $d^6$ (and even $d^7$) valence and substantial spin fluctuations, particularly into states which minimize double occupancy like those depicted in **f** and **g**. There is a slight preference to occupy $e_g$ states when orbitals are doubly occupied as in the $d^6$ multiplets.

In summary, we report the synthesis of a tetragonal perovskite phase, BCT-BaCoO$_3$, at 15 GPa and 1200 °C. HR-STEM analysis reveals BCT-BaCoO$_3$ crystallizes in the *I*4/*mcm* (#140) space group, adopting a EuTiO$_3$-type structure. No detectable oxygen vacancies were found in the structure, aligning with the Co$^{4+}$ characterization in XPS results. Magnetization measurements indicate ferromagnetic ordering at approximately 107 K and Curie-Weiss paramagnetic behavior above this temperature, alongside an LS-HS crossover upon cooling and a potential intermediate-spin state at high temperatures. Electrical resistivity experiments suggest insulating behavior with weak magnetoresistance. DFT and DFT+DMFT calculations suggest the insulating state originates from an orbitally selective transition sensitive to the Co-*d* shell's nominal valence. We conclude that BCT-BaCoO$_3$, exhibiting FM insulation below 107 K, offers a valuable platform for exploring spin-state transitions and unusual Co spin states in complex cobalt oxides.

## Methods

**High Pressure and High Temperature Synthesis.** The synthesis was conducted using a 1000-ton multi-anvil apparatus (MA) at the University of Michigan. The starting material was as-synthesized ambient pressure phase, which was prepared by thoroughly mixing the $BaCO_3$ and CoO in the atomic ratio of 1:1 and subsequently heating at 1100 °C for 72 hours before quenching.[31] The sample was kept at 110 °C overnight to remove the moisture before loading for high-pressure synthesis. The COMPRES 10/5 cell assemblies were used in the synthesis.[32] The sample was loaded in a platinum capsule and kept at 15 GPa and 1200 °C for 3 hours before quenching to room temperature and then decompressed to ambient pressure.

**Phase Analysis.** The phase identity and purity were examined using a Bruker Davinci powder X-ray diffractometer with Cu $K_\alpha$ radiation ($\lambda$ = 1.5406 Å). The upper and lower discriminator values were set to 0.40 V and 0.18 V, respectively, to mitigate the background due to fluorescence. Room temperature measurements were carefully performed with a step size of 0.020° at a scan speed of 5.00 sec/step over a Bragg angle ($2\theta$) range of 5–120°. The synthesis product was examined using the JEOL-7800FLV field emission SEM at the Robert B. Mitchell Electron Microbeam Analysis Lab (EMAL) of University of Michigan, and the analyses confirmed chemical purity and homogeneity.

**Structure and Chemical Composition Determination.** The structural and chemical composition determination was conducted using high-angle annular dark field (HAADF) imaging and energy dispersive X-ray spectroscopy (EDS) inside a scanning transmission electron microscope (STEM). The TEM sample was prepared by using a dual beam Helios instrument, which combines focused ion beam (FIB) and scanning electron microscopy. First, a cross-sectional lamella was extracted from the polycrystalline sample using FIB milling. The lamella gradually thinned down to approximately 200 nm at 30 kV. Subsequently, the sample was further reduced to thickness to around 50 nm at 5 kV. The final polishing of the sample was carried out at 2 kV. For HAADF imaging, a Themis-Z STEM microscope equipped with an aberration corrector and a monochromator was used. An acceleration voltage of 300 kV and a prove current of approximately 30 pA were employed for both STEM imaging and EDS mapping. In HAADF imaging, a convergence angle of 30 mrad and collection angle of 60 to 180 mrad were used.

**Chemical Valence State Analysis.** X-ray photoelectron spectroscopy (XPS) measurements were performed using a Thermo Fisher NEXSA spectrometer with a 125 mm mean radius, full 180° hemispherical analyzer, and 128-channel detector. This system uses a focused monochromatic Al $K_\alpha$ X-ray (1486.7 eV) source for excitation and an electron emission angle of 60 degrees. The narrow scan spectra were collected using a pass-energy of 50 eV with a step size of 0.1 eV. For the Ag $3d_{5/2}$ line, these conditions produced a FWHM of 0.84 eV ± 0.02 eV. The binding energy (BE) scale is calibrated using the Cu $2p_{3/2}$ feature at 932.62 ± 0.05 eV and Au $4f_{7/2}$ at 83.96 ± 0.05 eV.

**Physical Properties Measurement.** Magnetization measurements were carried out using a Quantum Design MPMS 3 magnetometer after demagnetization. Temperature dependent magnetization was measured over the temperature range of 2–300 K employing the zero-field cooled (ZFC) and field cooled (FC) protocols. Magnetic hysteresis loops were recorded with applied fields up to 7 T. Electrical resistivity measurements were conducted with four-probe method using platinum wires on a polycrystalline sample of BCT-BaCoO$_3$ in the dimensions of 1.0 × 0.8 × 1.0 mm$^3$ with a Quantum Design physical property measurement system (PPMS) DynaCool.

**Electronic Structure Calculation.** We conduct all-electron density functional theory (DFT) and charge self-consistent DFT with dynamical mean field theory (DFT+DMFT) of BCT-BaCoO$_3$ using *Portobello*[33-35]. The DFT equations are solved within the generalized-gradient approximation (GGA) using the Perdew-Burke-Ernzerhof (PBE)[36] functional, neglecting the spin-orbit coupling. An 8 × 8 × 8 $k$-mesh and basis with RK of 8 are used for all calculations. The DMFT equations are used to treat correlations within the Co $d$-shell, where the off-diagonal elements in the Hamiltonian are truncated in order to avoid a sign problem during the solution of the quantum impurity problem. We use a spherically symmetric Slater-Condon interaction[37] with Hubbard $U$ = 10 eV and Hund $J$ = 1 eV to describe the interaction, and we use the fully localized limit double-counting with an electron occupancy of $N_0$ = 5, which correspond to the nominal $d^5$ valence.

**X-ray Absorption Spectroscopy:** X-ray absorption spectroscopy (XAS) experiments were carried out at the REIXS beamline of the Canadian Light Source.[38] The experiment was carried out at normal incidence, at several temperatures between 20 K and 300 K, spanning the different regions identified from the Curie-Weiss fit of the magnetic susceptibility. Total electron yield was obtained from by measuring the drain current from the sample. The samples were mounted on silver paint to improve thermal contact. Data collection was carried out between 750 eV and 830 eV, spanning the Co $L_{2,3}$ and Ba $M_{4,5}$ resonances. Due to the overlap between the Co and Ba transitions, a BaTiO3 thin film was used as a $Ba^{2+}$ reference. The $BaTiO_3$ spectrum was used to remove the Ba M 4,5 white line from the $BaCoO_3$ sample.

## Acknowledgments

H.W., M.X. and W.X. at Michigan State University were supported by the U.S. Department of Energy (DOE), Division of Basic Energy Sciences under Contract DE-SC0023648. C.P. at MSU was supported by NSF grant DMR-2422361. The work at Princeton University was funded by the US Department of Energy, grant DE-FG02-98ER45796. STEM and XPS measurements along with the data analysis were supported by the U.S. DOE, Office of Science, Basic Energy Sciences, Division of Materials Sciences and Engineering, Synthesis and Processing Science Program under Award no. 10122. XPS was performed in the Environmental Molecular Sciences Laboratory (EMSL), a national scientific user facility sponsored by the DOE's Office of Biological and Environmental Research and located at PNNL. The work at Rutgers and BNL is supported by the US Department of Energy Office of Science Basic Energy Sciences as part of the Computational Materials Science Program. J.L. acknowledges support by NSF grant DMR-2422362. X.K. acknowledges the financial support by the U.S. Department of Energy, Office of Science, Office of Basic Energy Sciences, Materials Sciences and Engineering Division under DE-SC0019259. A.M.H. was supported by the Natural Sciences and Engineering Research Council of Canada (NSERC), the CIFAR Quantum Materials Program, and the Alfred P. Sloan Foundation. Part of the research described in this paper was performed at the Canadian Light Source, a national research facility of the University of Saskatchewan, which is supported by the Canada Foundation for Innovation (CFI), the Natural Sciences and Engineering Research Council (NSERC), the National Research Council (NRC), the Canadian Institutes of Health Research (CIHR), the Government of Saskatchewan, and the University of Saskatchewan.

# Supplementary Information

# Tetragonal BaCoO$_3$: A Co$^{4+}$ Ferromagnetic Mott Insulator with Inverted Spin Crossover

Mingyu Xu[1], Haozhe Wang[1], Krishna Prasad Koirala[2], Corey Melnick[3], Cheng Peng[1], Mario U. González-Rivas[4,5], Jiaqi Lu[6], Le Wang[2], Mark H. Engelhard[7], Yingge Du[2], Xianglin Ke[8], Robert J. Green[5,9], Alannah M. Hallas[4,5], Jie Li[6], Gabriel Kotliar[3,11], Weiwei Xie[1*]

11. Department of Chemistry, Michigan State University, East Lansing, MI 48824, USA
12. Physical and Computational Sciences Directorate, Pacific Northwest National Laboratory, Richland, WA 99354, USA
13. Condensed Matter Physics and Materials Science Department, Brookhaven National Laboratory, Upton, NY 11973, USA
14. Department of Physics & Astronomy, University of British Columbia, Vancouver, BC V6T 1Z1, Canada
15. Stewart Blusson Quantum Matter Institute, University of British Columbia, Vancouver, BC V6T 1Z4, Canada
16. Department of Earth and Environmental Sciences, University of Michigan, Ann Arbor, MI 48109, USA
17. Energy and Environment Directorate, Pacific Northwest National Laboratory, Richland, WA 99354, USA
18. Department of Physics and Astronomy, Michigan State University, East Lansing, MI 48824, USA
19. Department of Physics and Engineering Physics, University of Saskatchewan, Saskatoon S7N 5E2, Saskatchewan, Canada
20. Department of Physics and Astronomy, Rutgers University, Piscataway, NJ 08854, USA

*Corresponding authors: Dr. Weiwei Xie (xieweiwe@msu.edu)

**Supplementary Figures**

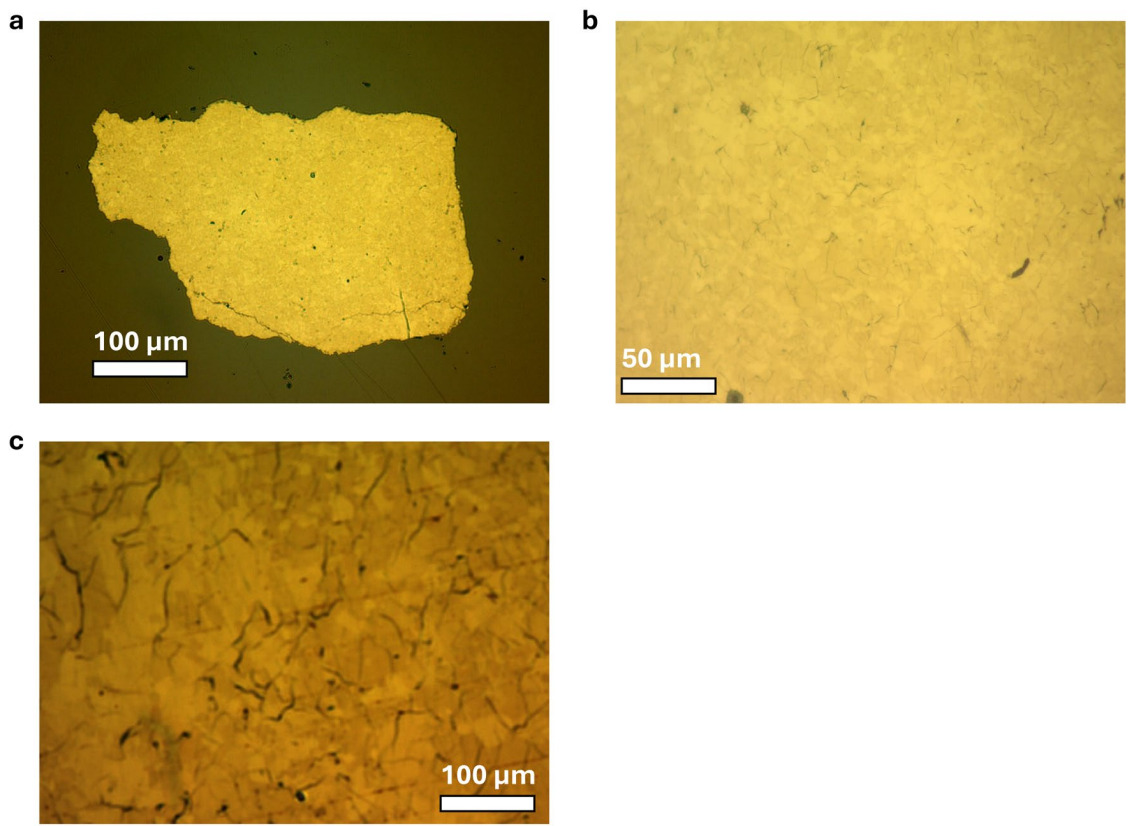

**Fig. S1 | Optical microscope images of t*I*-BaCoO₃. a,** Overall view of polished t*I*-BaCoO₃ with 5× magnification. **b,** Zoom-in view with 20× magnification. **c,** Zoom-in view in the contrast mode with 50× magnification.

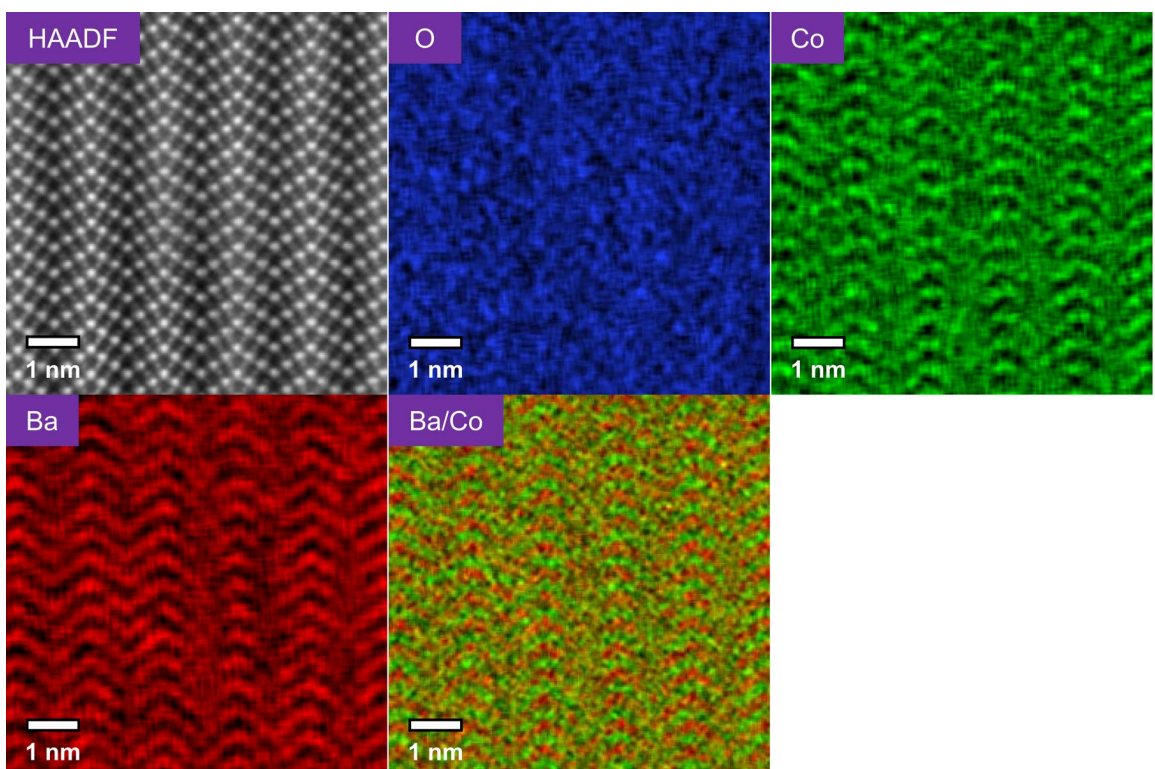

**Fig. S2 | Chemical element distribution of t*I*-BaCoO₃.** HAADF-STEM image and EDS mapping images of O, Co, Ba and Ba/Co are presented respectively.

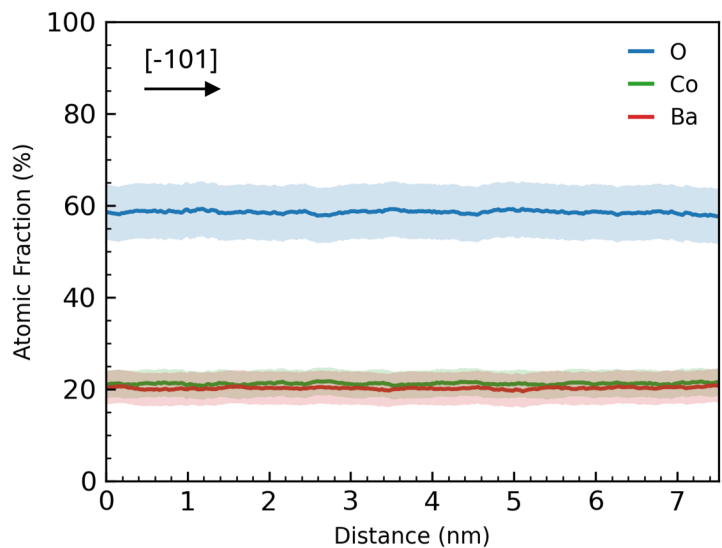

**Fig. S3 | Spatial variation of Ba, Co, and O based on EDS analysis.** The error bar is indicated by color filling.

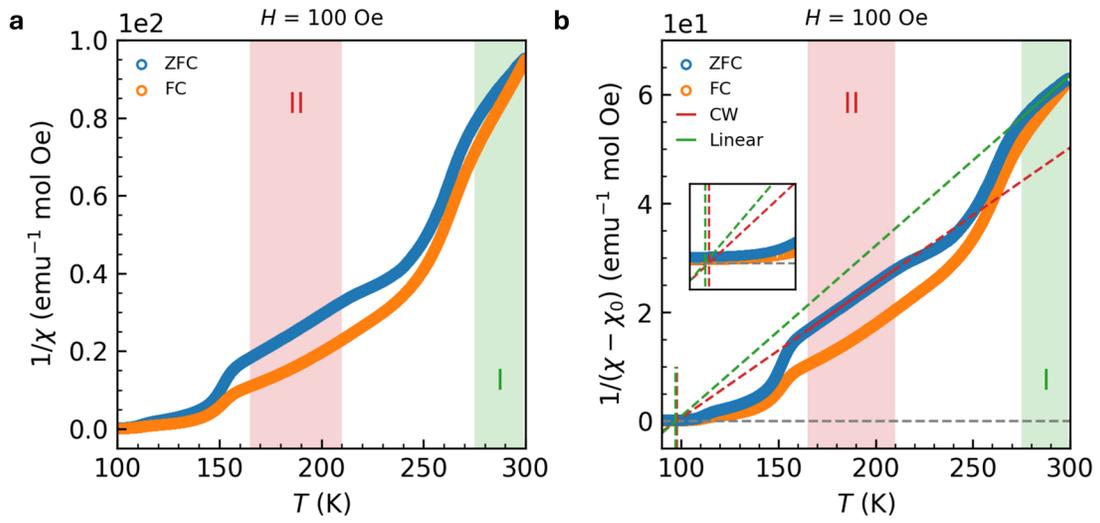

**Fig. S4 | Temperature-dependent magnetization of t*I*-BaCoO₃ in the range of 100–300 K. a,** Inverse magnetic susceptibility in temperature Regions I (green) and II (red). **b,** Magnetic susceptibility in temperature Region II fitted by Curie-Weiss law (red), in which Region I was linear fitted with the given $\chi_0$ (green). Inset, Zoom-in plot near 100 K.

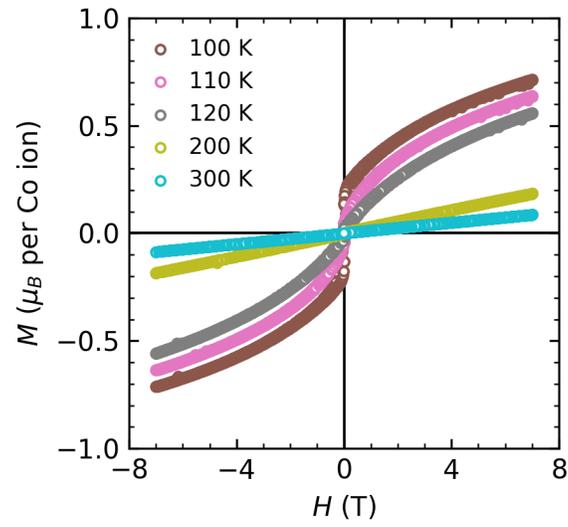

**Fig. S5 | Field dependent magnetization of t*I*-BaCoO₃ up to 7 T above 100 K.** Linear response to fields at high temperatures confirms its paramagnetic behavior.

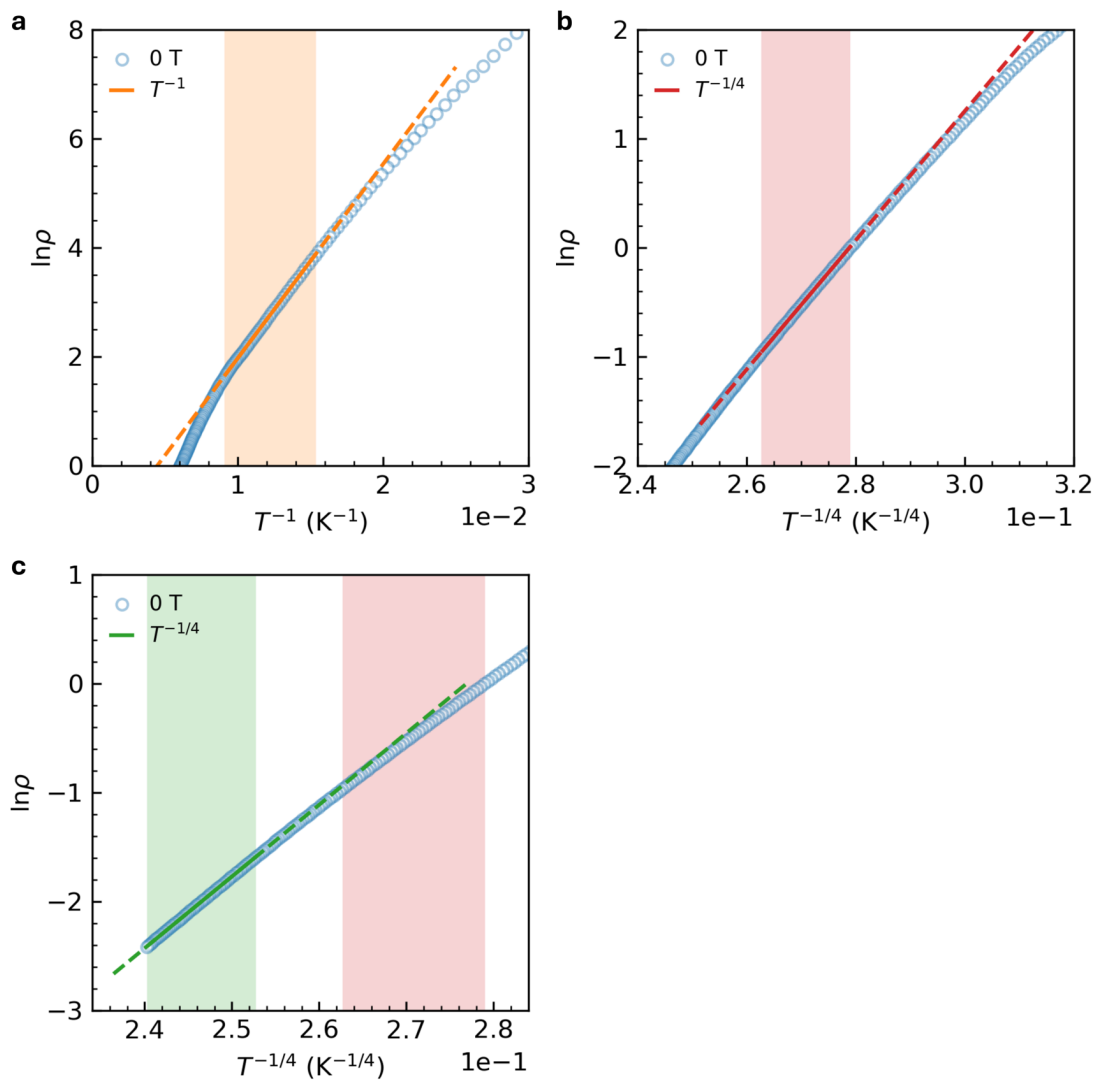

**Fig. S6 | Temperature variations of the zero-field electrical resistance. a,** ln $\rho$ versus $T^{-1}$ plot. **b-c,** ln $\rho$ versus $T^{-1/4}$ plot in Region I (green) and Region II (red). Temperature range of data fitting is indicated by color filling.

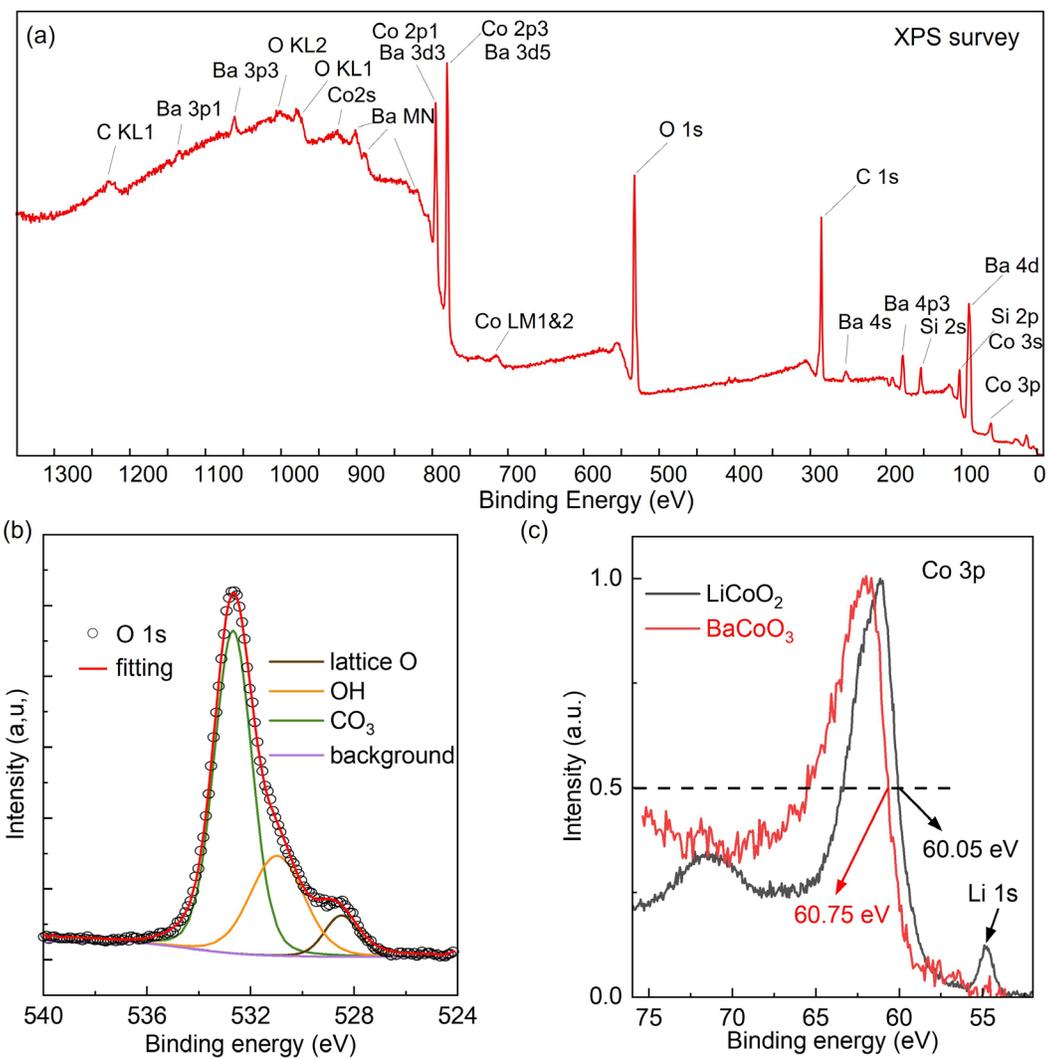

**Fig. S7 | Electronic structure and chemical valence analysis revealed by XPS. a,** Survey XPS spectra. **b-c,** O 1*s* and Co 3*p* core level XPS spectra.

# Supplementary Tables

**Table S1 | Spatial variation of Ba, Co, and O based on EDS analysis.** The values are averaged of 0-7.5 nm along [-101] direction.

| Element | Atomic Fraction / % | Atomic Fraction Error / % | Normalized Ratio |
|---------|---------------------|---------------------------|------------------|
| O       | 58.67               | 6.19                      | 2.78             |
| Co      | 21.14               | 3.60                      | 1                |
| Ba      | 20.09               | 3.04                      | 0.95             |

**Table S2 | Curie-Weiss fitting parameters of temperature dependent magnetic susceptibility.** The equation is presented below. $r^2$ is the coefficient of determination.

$$\frac{1}{\chi - \chi_0} = \frac{T - \theta_{CW}}{C}$$

where $\theta_{cw}$ is the paramagnetic Curie temperature, $\chi_0$ is the temperature independent susceptibility and $C$ is the Curie constant.

| | Curie-Weiss fitting | | | |
|---|---|---|---|---|
| Temperature Range / K | $\theta_{cw}$ / K | $C$ / (emu K/mol-Oe) | $\mu_{eff}$ / $\mu_B$ | $r^2$ |
| 275–299 | 225.6(4) | 0.3293(44) | 1.62(1) | 0.99984 |
| 165–210 | 97.8(2) | 4.025(1) | 5.67(1) | 0.99997 |

| Temperature Range / K | $\chi_0$ / (emu/mol-Oe) |
|---|---|
| 275–299 | 6.037(36) × 10$^{-3}$ |
| 165–210 | -5.370(74) × 10$^{-3}$ |

| | Linear | | | |
|---|---|---|---|---|
| Temperature Range / K | $\theta_{cw}$ / K | $C$ / (emu K /mol-Oe) | $\mu_{eff}$ / $\mu_B$ | $r^2$ |
| 275–299 | 97.2 | 3.191 | 5.05 | 0.99509 |
| 165–210 | 120.2 | 2.229 | 4.22 | 0.99803 |

**Table S3 | Thermal activation fitting parameters of electrical resistance.** The equation is presented below. $r^2$ is the coefficient of determination.

$$\ln \rho = \ln \rho_0 + \left(\frac{T_0}{T}\right)^\nu$$

where $\rho_0$ is the residual resistivity, and $T_0$ is the characteristic temperature.

| | $\nu = 1$ | | |
|---|---|---|---|
| Temperature Range / K | $\rho_0$ / ($\Omega$ cm$^{-1}$) | $T_0$ / K | $r^2$ |
| 65–110 | 0.2034 | 356.4 | 0.99996 |

| | $\nu = 1/4$ | | |
|---|---|---|---|
| Temperature Range / K | $\rho_0$ / ($\Omega$ cm$^{-1}$) | $T_0$ / K | $r^2$ |
| 165–210 | $6.623 \times 10^{-8}$ | $1.235 \times 10^7$ | 0.99993 |
| 245–300 | $1.217 \times 10^{-8}$ | $1.877 \times 10^7$ | 1.00000 |

## Supplementary Notes

**Note S1 | Chemical valence analysis of t*I*-BaCoO₃ revealed by XPS.**

To investigate the Co valence state in t*I*-BaCoO$_3$, X-ray photoelectron spectroscopy (XPS) experiments were carried out. A survey scan XPS spectrum of t*I*-BaCoO$_3$, covering the binding energy range of 0–1350 eV, is presented in **Fig. S7a**. The spectrum exhibits all characteristic lines corresponding to Ba, Co, and O, indicating the presence of these elements in the t*I*-BaCoO$_3$ sample. Additionally, a minor signal from Si was observed, which is attributed to residual contamination from the crucible used during the sample synthesis process. The Ba and Co ratio was estimated by analyzing the Ba 4*p* and Co 3*p* regions, yielding a ratio that is close to 1.

The O 1*s* spectrum (**Fig. S7b**) was analyzed by fitting with three distinct components. The low binding energy feature (528.5 eV) corresponds well with values reported in literature for bulk lattice oxygen.[1-5] At binding energy of 531.0 eV, a feature is attributed to the presence of hydroxyl group (OH). The most intense peak at 532.7 eV is indicative of the existence of carbonate species, which is supported by the significant contribution of the C 1*s* peak observed in the survey scan. Due to strong overlap between Ba 3*d* and Co 2*p* regions, the Co valence state was determined using the Co 3*p* spectrum (**Fig. S7c**). A Co$^{3+}$ reference sample of as-grown LiCoO$_2$ thin film on SrTiO$_3$ substrate was chosen for comparison.[6,7] To easily visualize the change in Co 3*p* line shape, we align these two Co 3*p* XPS spectra to place the corresponding O 1*s* (lattice O) peaks at 530.0 eV. Compared with the Co$^{3+}$ reference, Co 3*p* feature of t*I*-BaCoO$_3$ appeared broader and lacked an obvious satellite feature (~72 eV).[8] Moreover, a significant shift towards higher binding energy was observed in the Co 3*p* peak of t*I*-BaCoO$_3$. This shift, approximately 0.7 eV, indicates that the Co valence state in t*I*-BaCoO$_3$ is considerably higher than Co$^{3+}$, close to Co$^{4+}$.[9,10]